\documentclass[aps,prb,preprint]{revtex4-1}
\usepackage{amsmath,amssymb}
\usepackage{bm}
\usepackage{epsfig}

\renewcommand{\d}{\text{d}}

\newcommand{\nl}{\nonumber \\}
\newcommand{\la}{\langle}
\newcommand{\ra}{\rangle}
\newcommand{\lla}{\la\!\la}
\newcommand{\rra}{\ra\!\ra}

\newcommand{\greater}{\mbox{\tiny $>$}}
\newcommand{\lesser}{\mbox{\tiny $<$}}

\newcommand{\Sch}{Schr\"{o}dinger\ }
\newcommand{\Hei}{Heisenberg\ }

\newcommand{\Sec}[1]{Sec.\,\ref{#1}}

\newcommand{\be}{\begin{equation}}
\newcommand{\ee}{\end{equation}}
\newcommand{\bea}{\begin{eqnarray}}
\newcommand{\eea}{\end{eqnarray}}
\newcommand{\bsube}{\begin{subequations}}
\newcommand{\esube}{\end{subequations}}
\newcommand{\Eq}[1]{Eq.\,(\ref{#1})}
\newcommand{\Eqs}[1]{Eqs.\,(\ref{#1})}
\newcommand{\Fig}[1]{Fig.\,\ref{#1}}

\newcommand{\B}{\mbox{\tiny B}}
\newcommand{\D}{\mbox{\tiny D}}

\newcommand{\T}{\mbox{\tiny T}}
\newcommand{\tS}{\mbox{\tiny S}}

\newcommand{\BO}{\mbox{\tiny BO}}
\newcommand{\SB}{\mbox{\tiny SB}}

\newcommand{\leftact}{\overset{\rightarrow}}
\newcommand{\rightact}{\overset{\leftarrow}}

\newcommand{\iChEM}{{\it i}{\rm ChEM}}

\begin{document}

\title{Effects of Herzberg--Teller vibronic coupling on coherent excitation energy transfer}

\author{Hou-Dao Zhang}\email{hdz@ustc.edu.cn}
\author{Qin Qiao}
\author{Rui-Xue Xu}
\author{YiJing Yan}\email{yanyj@ustc.edu.cn}

\affiliation{
Hefei National Laboratory for Physical Sciences at the Microscale
and Department of Chemical Physics
and Synergetic Innovation Center of Quantum Information and Quantum Physics
and \iChEM,
University of Science and Technology of China, Hefei, Anhui 230026, China
}

\begin{abstract}
 In this work, we study the effects of non-Condon vibronic coupling
on the quantum coherence of excitation energy transfer,
via the exact dissipaton-equation-of-motion (DEOM) evaluations on excitonic model systems.
Field-triggered excitation energy transfer dynamics
and two dimensional coherent spectroscopy are simulated
for both Condon and non-Condon vibronic couplings.
Our results clearly demonstrate that the non-Condon vibronic coupling
intensifies the dynamical electronic-vibrational energy transfer
and enhances the total system-and-bath quantum coherence.
Moreover, the hybrid bath dynamics for
non-Condon effects enriches the theoretical calculation, and further sheds
light on the interpretation of the experimental nonlinear spectroscopy.
\end{abstract}

\maketitle

\section{Introduction}
\label{thsec1}
The long-lived quantum beats in the coherent excitation energy transfer (EET) in
biological light-harvesting systems have drawn plentiful attention
in the past decade.\cite{Eng07782,Che084254,Pan1012766,Ple13235102}
Two dimensional (2D) spectroscopy studies indicate that
both electronic and vibrational coherence
and their entanglement are responsible for this phenomenon.
\cite{Tur111904,Tem1412865,But14034306,Cam1595,Bal1519491}
This interpretation is confirmed by many theoretical works 
in the framework of the Franck--Condon approximation.\cite{Man121497,Per14404,Fuj15212403,Mon15065101}
In the meanwhile, however, the non-Condon vibronic coupling is also proposed to
enhance the quantum coherence in many organic and biological molecular systems.\cite{Ish082219, Kam0010637, Wom12154016}
To obtain a better understanding of the origin of quantum coherence in EET dynamics,
the interplay of the Condon and non-Condon vibronic coupling effects
deserves more theoretical investigations.

In studying the vibronic coupling effects, 
we need an integrated theory
to handle both the excitonic system and the vibrational environment.
On the other hand,
the nonlinear nature of 2D spectroscopy calls for non-perturbative and
non-Markovian treatments, based on such as
the hierarchical equations of motion (HEOM) approach.\cite{Tan89101,Yan04216,Ish053131,Tan06082001,Tan14044114,Tan15144110,Xu05041103,Xu07031107,Zhe121129,Jin08234703,Che11194508}
However, HEOM focuses primarily on the reduced system only,
despite it involves a vast number of auxiliary density operators,
which are purely mathematical quantities.
As an alternative exact approach for open systems,
the recently developed dissipaton equation of motion (DEOM) theory
\cite{Yan14054105,Zha15024112,Jin15234108}
provides a quasi-particle dissipaton picture for the hybrid bath,
and identifies its dynamical quantities
as multi-body dissipaton density operators. In this manner, the DEOM
is capable of treating both system and hybrid bath dynamics non-perturbatively.
On describing reduced system dynamics,
the DEOM recovers the HEOM method.\cite{Tan89101,Yan04216,Ish053131,Tan06082001,Tan14044114,Tan15144110,Xu05041103,Xu07031107,Zhe121129,Jin08234703,Che11194508}
The optimized DEOM formalism with damped
Brownian oscillators can be constructed
for studying the system dynamics under the influence of solvent fluctuations
and intramolecular vibrations.\cite{Xu09214111,Zhu115678,Din11164107,Din12224103}
The DEOM further enables us to
directly investigate the solvation bath dynamics.
In fact, it has been successfully applied in studying
a variety of Fano-type interferences,\cite{Zha15024112,Zha15214112,Xu151816}
quantum transport current shot noise spectrum,\cite{Jin15234108}
and solvent-induced non-Condon polarization effects.\cite{Zha16CP}

In this work, we study the influence of 
non-Condon vibronic coupling on the coherent EET dynamics 
in excitonic model systems and compare it with the Condon counterpart,
via the DEOM approach.
In simulating the electronic excitation of the system,
the vibrational coordinates are explicitly involved in the transition dipole moments.
This accounts for the non-Condon vibronic coupling 
under the Herzberg-Teller approximation. \cite{Orl734513,Lin743802,Tan932263, Tan93118,Ish082219}
We model these vibrational modes in terms of underdamped Brownian oscillators (BOs).\cite{Tho012991,Zhu115678,Din12224103}
The underlying non-Condon effects are therefore studied via the hybrid bath dynamics 
in the DEOM framework.\cite{Zha16CP}
In the Franck--Condon approximation, however, the transition dipole moments
are independent of nuclear coordinates and
the underdamped BOs merely serve as the vibrational environment.
Based on our simulations on field-triggered EET dynamics
and coherent 2D spectroscopy,
we conclude that the non-Condon vibronic coupling facilitates the
dynamical electronic-vibrational energy transfer and enhances
the total system-and-bath quantum coherence, at both cryogenic and
room temperatures. 
While both the Condon and non-Condon vibronic coupling elongate the quantum beats, 
the non-Condon polarization improves the quantum yield
of the excitonic system.

This paper is organized as follows.
In \Sec{thsec2}, we first set up an excitonic multi-chromophore system
and model its surrounding environment with multiple Brownian oscillators,
then briefly review the DEOM method
and exemplify its application on the hybrid bath dynamics
for non-Condon polarization effects.
In \Sec{thsec3}, we present
the simulation results of field-triggered EET dynamics and coherent 2D spectroscopy
for excitonic dimer systems,
and compare the non-Condon and Condon vibronic coupling effects.
The temperature dependence of these two effects is also discussed.
Finally, we conclude this work in \Sec{thconc}.

Throughout this paper, we set $\hbar=1$ and $\beta=1/(k_BT)$,
with $k_B$ and $T$ being the Boltzmann constant
and temperature, respectively.

\section{Methodology}
\label{thsec2}
\subsection{Model Setup}
\label{thsec2A}
Consider an excitonic multi-chromophore system embedded in an equilibrated thermal bath.
The total composite system-and-bath Hamiltonian is summarized as
$H_{\T}=H_{\tS}+H_{\SB}+h_{\B}$.
The excitonic system Hamiltonian reads
\be
  H_{\tS}=\epsilon_g|g\ra\la g|+\sum_a\epsilon_a|a\ra\la a|+\sum_{a\ne b}V_{ab}|a\ra\la b|.
\ee
Here, the ground state energy $\epsilon_g$ is often set to be zero,
whereas the diagonal onsite energies and inter-site couplings
denote $\{\epsilon_a\}$ and $\{V_{ab}\}$.
The electronic dipole moment of each chromophore is
$\hat\mu_a=\mu_a({\bf X}_a)(|g\ra\la a|+|a\ra\la g|)$,
where the transition matrix element $\mu_a({\bf X}_a)$
depends on the nuclear coordinates ${\bf X}_a$.
This form in fact accounts for the non-Condon coupling
under the Herzberg--Teller approximation.\cite{Lin743802}
In the Franck--Condon limit,
$\mu_a({\bf X}_a)$ becomes a constant without dependence on ${\bf X}_a$.

The thermal bath is modeled as numerous harmonic oscillators,
$h_{\B}=\sum_j\omega_j(p_j^2+x_j^2)/2$, and
the system-bath coupling takes the factorization form, i.e.,
\be
  H_{\SB}=\sum_a\hat Q_a\hat F_a,
\ee
where $\hat Q_a=|a\ra\la a|$ is a system operator
and $\hat F_a=\sum_jc_{aj}x_j$ is a linear collective bath operator.
The bath influence on the reduced system
can be entirely characterized by the hybrid bath spectral density
$J_{ab}(\omega)=\frac{\pi}{2}\sum_jc_{aj}^2\delta(\omega-\omega_j)$,
or the macroscopic bare-bath correlation function (setting $t>0$ hereafter),
\be\label{Ct}
  \la\hat F_a(t)\hat F_b(0)\ra_{\B}
=\frac{1}{\pi}\int_{-\infty}^{\infty}\!d\omega\,
 \frac{e^{-i\omega t}J_{ab}(\omega)}{1-e^{-\beta\omega}}
\approx\sum_k\eta_{abk}e^{-\gamma_kt},
\ee
with $\hat F_a(t)=e^{ih_{\B}t}\hat F_ae^{-ih_{\B}t}$ being
a Gaussian stochastic variable in the $h_{\B}$-interaction picture.
The first identity is the fluctuation--dissipation theorem,
\cite{Wei08,Yan05187}
while the second involves the parameterization of $J_{ab}(\omega)$
and certain sum-over-pole expansion of Bose function,
$f^{\rm Bose}(\omega)=(1-e^{-\beta\omega})^{-1}$,
followed by the Cauchy contour integral.
The time-reversal correspondence to \Eq{Ct} reads
\be\label{Ctr}
  \la\hat F_b(0)\hat F_a(t)\ra_{\B}
=\sum_k\eta_{ab\bar k}^{\ast}e^{-\gamma_kt}.
\ee
The second expression, where the associate index
$\bar k$ goes with
$\gamma_{\bar k}=\gamma_k^{\ast}$,
reflects the fact that
$\gamma_k$ and $\gamma_k^{\ast}$ must both appear if they
are complex.
Apparently, $\bar k=k$ for a real $\gamma_k$.

For the bath spectral density, 
we adopt the multiple Brownian oscillators (BOs) model.
It can describe optically active vibronic coupling 
via the underdamped BO mode and 
also energy fluctuation via strongly overdamped Drude dissipation.
For simplicity, we omit the subscript of different dissipative modes
to the end of this subsection.
An individual BO spectral density reads\cite{Muk95}
\be\label{JBO}
  J_{\BO}(\omega)=
  \frac{2\lambda_{\BO}\zeta_{\BO}\omega_{\BO}^2\omega}{(\omega^2-\omega_{\BO}^2)^2+\zeta_{\BO}^2\omega^2},
\ee
with $\lambda_{\BO}$, $\omega_{\BO}$ and $\zeta_{\BO}$
being the reorganization energy, vibrational frequency and
the phonon linewidth, respectively.
The Huang--Rhys factor that reflects the exciton-phonon coupling strength
is defined as $S=\lambda_{\BO}/\omega_{\BO}$ in the free-oscillator limit.
The underdamped BO is characterized by $\zeta_{\BO}/2<\omega_{\BO}$,
and is often adopted to describe intramolecular vibrations.
In the strongly overdamped region ($\zeta_{\BO}/2\gg\omega_{\BO}$),
\Eq{JBO} recovers the Drude form, \cite{Tho012991,Din12224103}
\be\label{JD}
  J_{\D}(\omega)=
  \frac{2\lambda_{\D}\gamma_{\D}\omega}{\omega^2+\gamma_{\D}^2},
\ee
with $\lambda_{\D}=\lambda_{\BO}$ and $\gamma_{\D}=\omega_{\BO}^2/\zeta_{\BO}$.
This spectral density is often used to model diffusive solvent fluctuations.
\cite{Din11164107,Zhu115678}

\subsection{The DEOM theory}
\label{thsec2B}

In DEOM, the dissipaton decomposition of the hybridization bath operator $\hat F_a$ follows
\be\label{Fdecom}
  \hat F_{a}=\sum_{k}\hat f_{ak}.
\ee
Each dissipaton $\hat f_{ak}$ corresponds to
an individual exponential term in \Eq{Ct} or \Eq{Ctr}, described by
\be\label{ff}
\begin{split}
 &\la\hat f_{ak}(t)\hat f_{bj}(0)\ra_{\B}^{\greater}
=\delta_{kj}\eta_{abk}e^{-\gamma_kt},
\\
 &\la\hat f_{bj}(0)\hat f_{ak}(t)\ra_{\B}^{\lesser}
=\delta_{kj}\eta_{ab\bar k}^{\ast} e^{-\gamma_kt}.
\end{split}
\ee
The same damping exponent along the time forward ($\greater$) and backward ($\lesser$) pathways
highlights the diffusive nature of dissipatons.\cite{Yan14054105,Zha15024112}
The above dissipaton description of bath correlation is mathematically valid,
and can be further simplified with $\bar k=k$
for individual real-colored dissipaton
in the Smoluchowski diffusive limit.
For an underdamped BO,
there exists a pair of dissipatons with complex conjugate
damping exponents, $\gamma_{\bar k}=\gamma_k^{\ast}$.
In this case, a more physically comprehensive interpretation
that involves second quantization treatment can be put forward,
which neatly reflects the quasi-particle nature of dissipatons.
Detailed derivation and interpretation on this issue
will be published elsewhere.

Dynamical variables of DEOM are the so-called
\emph{dissipaton density operators} (DDOs),
defined as\cite{Yan14054105,Zha15024112}
\be\label{DDO_def}
  \rho_{\bf n}^{(n)}(t)\equiv
  \text{tr}_{\B}\Big[\Big(\prod_{ak}\hat f_{ak}^{n_{ak}}\Big)^{\circ}\rho_{\T}(t)\Big].
\ee
Here, $\rho_{\T}(t)$ is the total system-and-bath composite density operator,
and ${\rm tr}_{\B}$ denotes the trace over bath degrees of freedom.
The dissipaton operators product inside the circled parentheses is \emph{irreducible}.
The notation here follows $(\text{c-number})^{\circ}=0$,
and bosonic dissipatons are permutable,
$\big(\hat f_{ak}\hat f_{bj}\big)^{\circ}
=\big(\hat f_{bj}\hat f_{ak}\big)^{\circ}$.
In \Eq{DDO_def}, ${\bf n}\equiv\{n_{ak}\}$ and $n\equiv\sum_{ak}n_{ak}$,
with $n_{ak}$ being the participation number of individual dissipaton.
Apparently, $\rho^{(0)}=\text{tr}_{\B}\rho_{\T}=\rho_{\tS}$
is the reduced system density operator.
Those $\{\rho_{\bf n}^{(n>0)}\}$ specify $n$-body dissipaton configurations,
which account for the hybrid bath dynamics.
Denote also the adjacent $(n\pm1)$-body quantities,
$\rho^{(n+1)}_{{\bf n}_{ak}^+}$
and
$\rho^{(n-1)}_{{\bf n}_{ak}^-}$,
with the index ${\bf n}_{ak}^{\pm}$ differing from ${\bf n}$
by $n_{ak}\pm1$, at the specified dissipaton.
The multi-body DDOs form the hierarchy structure
and they are dynamically linked together
via the irreducible notation in \Eq{DDO_def} and the generalized Wick's theorem \cite{Yan14054105,Zha15024112}:
\begin{align}\label{Wick_greater}
&\quad\,
 {\rm tr}_{\B}\Big[\Big(\prod_{ak}\hat f^{n_{ak}}_{ak}\Big)^{\circ}
  \hat f_{bj}\rho_{\T}\Big]
\nl&
 =\rho^{(n+1)}_{{\bf n}^{+}_{bj}}
+\sum_{ak} n_{ak} \la\hat f_{ak}\hat f_{bj}\ra^{\greater}_{\B}
  \rho^{(n-1)}_{{\bf n}^{-}_{ak}},
\nl&
 =\rho^{(n+1)}_{{\bf n}^{+}_{bj}}
+\sum_a n_{aj}\eta_{abj}
  \rho^{(n-1)}_{{\bf n}^{-}_{aj}},
\end{align}
and
\begin{align}\label{Wick_lesser}
&\quad\,
 {\rm tr}_{\B}\Big[\Big(\prod_{ak}\hat f^{n_{ak}}_{ak}\Big)^{\circ}
  \rho_{\T}\hat f_{bj}\Big]
\nl&=\rho^{(n+1)}_{{\bf n}^{+}_{bj}}
 +\sum_{ak} n_{ak} \la\hat f_{bj}\hat f_{ak}\ra^{\lesser}_{\B}
  \rho^{(n-1)}_{{\bf n}^{-}_{ak}}.
\nl&=\rho^{(n+1)}_{{\bf n}^{+}_{bj}}
 +\sum_a n_{aj}\eta_{ab\bar j}^{\ast}
  \rho^{(n-1)}_{{\bf n}^{-}_{aj}},
\end{align}
where we use the following relation [cf.\ \Eq{ff}]
\begin{equation*}
\begin{split}
&\la \hat f_{ak}\hat f_{bj}\ra^{\greater}_{\B}
\equiv \la \hat f_{ak}(0+)\hat f_{bj}\ra
 = \delta_{kj}\eta_{abk},
\\
&\la \hat f_{bj}\hat f_{ak}\ra^{\lesser}_{\B}
\equiv \la \hat f_{bj}\hat f_{ak}(0+)\ra
=\delta_{kj}\eta^{\ast}_{ab\bar{k}}.
\end{split}
\end{equation*}
Equations (\ref{Wick_greater}) and (\ref{Wick_lesser})
accomplish the most significant part of the DEOM theory.
They are adopted not only in deriving the
fundamental DEOM formalism below,
but also for evaluating dynamics of hybrid bath quantities.
\cite{Yan14054105,Zha15024112,Zha16CP}

The DEOM formalism can be derived via the established dissipaton algebra.
\cite{Yan14054105,Zha15024112}
The final results read
\begin{align}\label{DEOM_sch}
\dot\rho^{(n)}_{\bf n}=&
-\big[i{\cal L}_{\tS}+\sum_{ak} n_{ak}\gamma_k\big]\rho^{(n)}_{\bf n}
\nl&
-i\sum_{ak}\Big[
  {\cal A}_a\rho^{(n+1)}_{{\bf n}_{ak}^+}+
  n_{ak}{\cal C}_{ak}\rho^{(n-1)}_{{\bf n}_{ak}^-}
\Big]\, ,
\end{align}
where ${\cal L}_{\tS}\hat O=[H_{\tS},\hat O]$
defines the system Liouvillian and
\be\label{calAC}
\begin{split}
{\cal A}_a\hat O&\equiv \big[\hat Q_a,\hat O\big],
\\
{\cal C}_{ak}\hat O&\equiv \sum_b\big(\eta_{abk}\hat Q_b\hat O
  -\eta_{ab{\bar k}}^{\ast}\hat O\hat Q_b\big).
\end{split}
\ee
Equation (\ref{DEOM_sch}) recovers the HEOM formalism,
\cite{Tan89101,Yan04216,Ish053131,Tan06082001,Tan14044114,Tan15144110,Xu05041103,Xu07031107,Zhe121129}
constructed originally on the basis of Feynman-Vernon
influence functional path integral expression \cite{Fey63118}.
Nevertheless, the DDOs are now responsible for hybrid bath dynamics
instead of being pure mathematical tools in HEOM. \cite{Tan89101,Yan04216,Ish053131,Tan06082001,Tan14044114,Tan15144110,Xu05041103,Xu07031107,Zhe121129}
This fact has been justified by many DEOM-based studies on Fano interference,
current shot noise, and solvent-induced non-Condon polarization phenomena.\cite{Zha15024112,Zha15214112,Jin15234108,Zha16CP}
In this work, we also illustrate the DEOM study of hybrid bath dynamics for
non-Condon vibronic coupling effects.

Denote $\bm{\rho}\equiv\{\rho^{(0)},\rho_{\bf n}^{(n)}\}$ the DEOM-space state vector.
Similarly, $\hat{\bm{A}}\equiv\{\hat A^{(0)},\hat A_{\bf n}^{(n)}\}$
is introduced as the DEOM-space extension of a dynamical variable,
$\hat A$, in the system-and-hybrid-bath subspace.
According to ${\rm Tr}(\dot{\hat A}\rho_{\T})={\rm Tr}(\hat A\dot\rho_{\T})$,
we immediately obtain the DEOM in the \Hei picture \cite{Xu11497,Zhu115678,Xu13024106},
\begin{align}\label{DEOM_hei}
\dot{\hat A}^{(n)}_{\bf n} &=
-\hat A^{(n)}_{\bf n}\big(i{\cal L}_{\tS}+\sum_{ak}n_{ak}\gamma_k\big)
\nl&\quad
-i\sum_{ak}\Big[\hat A^{(n-1)}_{{\bf n}_{ak}^-}{\cal A}_a
+(n_{ak}+1)\hat A^{(n+1)}_{{\bf n}_{ak}^+}{\cal C}_{ak}\Big].
\end{align}
The involving superoperators in Heisenberg picture read
$\hat O{\cal L}_{\tS}\equiv[\hat O,H_{\tS}]$ and
\be
\begin{split}
\hat O{\cal A}_a&\equiv \big[\hat O,\hat Q_{a}\big],
\\
\hat O{\cal C}_{ak}&\equiv \sum_b\big(\eta_{abk}\hat O\hat Q_{b}
-\eta_{ab\bar k}^{\ast}\hat Q_{b}\hat O\big).
\end{split}
\ee
The Heisenberg picture dynamics is equivalent to the \Sch picture
counterpart, since the expectation value of any physical observable
is of prescription invariance.
Furthermore, the mixed Heisenberg-\Sch picture dynamics
has been implemented in the DEOM/HEOM method,
for efficient evaluations of nonlinear correlation and response functions. \cite{Xu11497,Zha16CP}
In addition, an efficient and prescription invariance truncation scheme
is also developed for the DEOM/HEOM evolution
in both \Sch and Heisenberg pictures. \cite{Zha15214112,Zha16CP}

\subsection{The non-Condon dynamics via DEOM approach}
\label{thsec2C}

To exemplify the DEOM application in bath dynamics,
we study the non-Condon vibronic coupling effects in the linear absorption spectroscopy,
where the polarizable vibrational modes are treated in terms of
underdamped BOs in the hybrid bath.

Under the dipole approximation,
the matter-field interaction is given by
$H_{\text{mat-fld}}(t)=-\hat V\varepsilon(t)$,
with the external field being $\varepsilon(t)$, and the total dipole moment
\be
  \hat V=\sum_a\mu_a({\bf X}_a)\hat D_a.
\ee
Here, $\mu_a({\bf X}_a)$ and $\hat D_a=|g\ra\la a|+|a\ra\la g|$
are the transition matrix element and electronic operator, respectively.
The non-Condon vibronic coupling under the Herzberg--Teller approximation
involves the nuclear coordinate dependence in the dipole moment,
for which we consider the simple form
\be\label{mua}
  \mu_a({\bf X}_a)=\mu_a+\mu'_a\hat F_a^{\BO}/\lambda_{\BO},
\ee
where the hybrid bath operator $\hat F_a^{\BO}$ is associated with an underdamped BO.
In the Franck--Condon limit, the dipole moment element
is a constant, with $\mu_a({\bf X}_a)=\mu_a$.

The linear absorption lineshape is simulated via
\begin{align}\label{Sw}
  I(\omega)&={\rm Re}\,\int_0^{\infty}\!dt\,e^{i\omega t}
  {\rm Tr}[\hat Ve^{-i{\cal L}_{\T}t}\hat V\rho_{\T}^{\rm eq}]
\nl
 &={\rm Re}\,\int_0^{\infty}\!dt\,e^{i\omega t}
  \lla\hat{\bm{V}}|e^{-i\bm{{\cal L}}t}\hat{\bm{V}}|{\bm\rho}^{\rm eq}\rra.
\end{align}
The Fourier integrand in the first line defines the dipole-dipole correlation
in the total system-and-bath space,
with $\rho_{\T}^{\rm eq}$ being the equilibrated total density operator
and ${\cal L}_{\T}$ the total Liouvillian,
while the one in the second line is the DEOM-space equivalence.
In numerical implementation, the DEOM-space evaluation is carried out
following the established dissipaton algebra,\cite{Yan14054105,Zha15024112,Zha16CP}
including the dissipaton decomposition of
bath operator in \Eq{Fdecom}, the generalized Wick's theorem in \Eqs{Wick_greater}
and (\ref{Wick_lesser}), and the ordinary DEOM evolution via \Eq{DEOM_sch}.

\begin{figure}
\includegraphics[width=0.5\textwidth]{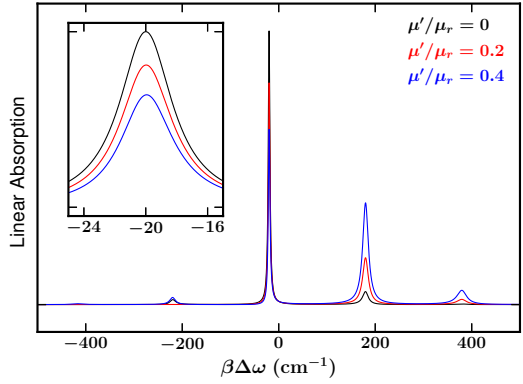}
\caption{
Linear absorption lineshapes of an excitonic monomer
under the influence of an underdamped BO at $298$\,K,
with various $\mu'$ and fixed $\mu=\mu_r$,
where $\mu_r$ is the reference dipole strength.
Parameters  of the underdamped BO are
$S=0.1$, $\omega_{\BO}=200\,{\rm cm}^{-1}$ and
$\zeta_{\BO}=10\,{\rm cm}^{-1}$.
}\label{fig1}
\end{figure}

Figure \ref{fig1} depicts the linear absorption results
for an excitonic monomer under the influence of
a bare underdamped BO mode.
In fact, for such two-level system,
the linear absorption profile is analytically solvable,
resulting in
\be\label{Sana}
I(\omega)=(\mu+\mu'\Delta\omega/\lambda_{\BO})^2I_0(\omega).
\ee
Here, $\Delta\omega=\omega-\epsilon$ where $\epsilon$ is the excitonic onsite energy, and
$I_0(\omega)={\rm Re}\,\int_0^{\infty}\!dt\,e^{i\Delta\omega t-g(t)}$
denotes the result in the Franck--Condon limit,
with $g(t)$ being the well-known lineshape function for the simple two-level system.\cite{Muk95}
We have verified that the analytical solutions are identical to the results in \Fig{fig1}.
In the Condon limit, which corresponds to $\mu'=0$,
we observe the sharp zero-phonon transition peak at $\Delta\omega=-20\,{\rm cm}^{-1}$,
and small vibrational signals about $\omega_{\B}$
away on both red and blue sides.
As $\mu'$ is tuned up, i.e., the non-Condon transition increases,
the zero-phonon transition peak decreases
and vibrational excitations become prominent.
The absorption lineshape would develop into two continuum bands
when BO goes to the super-overdamped region.
This is equivalent to the solvent-induced non-Condon polarization
effect, which has already been studied in our previous work.\cite{Zha16CP}

\begin{figure}
\includegraphics[width=0.5\textwidth]{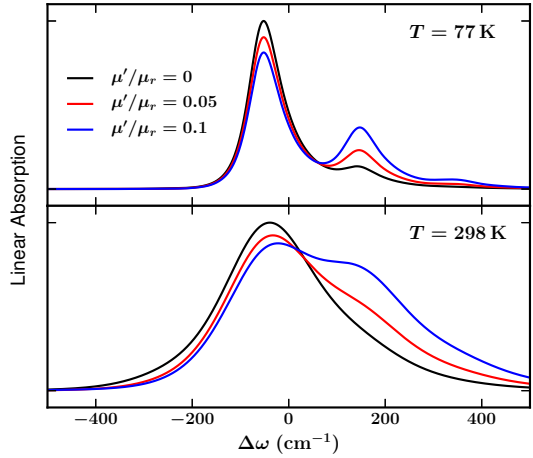}
\caption{
Linear absorption lineshapes of the excitonic monomer in \Fig{fig1}
with additional solvent environment at $77$ and $298$\,K,
for various non-Condon transition strengths.
In addition to the underdamped BO,
the solvent environment is modeled by a Drude spectral density,
with parameters $\lambda_{\D}=40\,{\rm cm}^{-1}$
and $\gamma_{\D}=100\,{\rm cm}^{-1}$.
}\label{fig2}
\end{figure}

Figure \ref{fig2} presents the linear absorption results
that additionally involve solvent environment described by Drude dissipation,
while all other parameters (except $\mu'$) remain the same as \Fig{fig1}.
Due to the solvent modulation,
both electronic and vibrational peaks
are now much broader than the ones in \Fig{fig1}, 
which only contains the vibronic coupling modeled by 
underdampled BO.
Apparently, the non-Condon effects are more prominent at the room temperature ($298$\,K).
At $77$\,K, the increase of $\mu'$ diminishes the zero-phonon transition
and strengthens the vibrational peaks on the blue side,
but does not change the peak positions significantly.
At $298$\,K, 
the excitonic peak becomes much broader and shifted to the blue side, while 
the vibrational one gradually appears, as the $\mu'$ increases.

\begin{figure}
\includegraphics[width=0.5\textwidth]{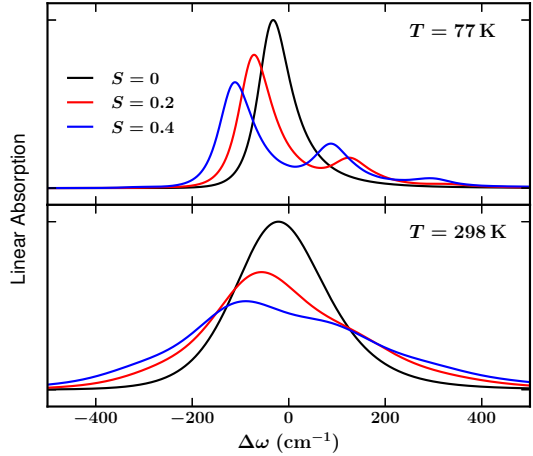}
\caption{
Linear absorption lineshapes of the excitonic monomer in
\Fig{fig2} in the Condon limit at $77$ and $298$\,K, for various
Huang-Rhys factors.
}\label{fig3}
\end{figure}

Last but not least, we note that the appearance of the vibrational peaks may 
be caused by not only the non-Condon transition,
but also the strong Franck--Condon vibronic coupling. 
In the latter case, the peaks are red shifted instead, as shown in \Fig{fig3}. 
To distinguish the above two causes experimentally, 
one can examine the mirror symmetry between the absorption and 
emission spectroscopies.\cite{Orl734513} 
While the Franck--Condon approximation results in symmetric lineshapes,
the non-Condon effects preferentially 
enhance the vibrational peaks on the blue side in both the absorption and emission spectroscopy.

\section{Results and Discussion}
\label{thsec3}

In this section, we discuss the simulation results of 
field-triggered EET dynamics and 2D spectroscopy 
for an excitonic dimer system.
The system and bath parameters are as follows.
The site energies are $\epsilon_1=12500$\,${\rm cm}^{-1}$ and $\epsilon_2=12300$\,${\rm cm}^{-1}$,
and the dipole-dipole coupling is $V_{12}=-100$\,${\rm cm}^{-1}$.
Each site is longitudinally subject to a diffusive solvent bath
and a vibrational mode,
which are characterized by a Drude model 
($\lambda_{\D}=40$\,${\rm cm}^{-1}$ and $\gamma_{\D}=100$\,${\rm cm}^{-1}$),
and an underdamped BO ($S=0.1$, $\omega_{\BO}=200$\,${\rm cm}^{-1}$
and $\zeta_{\BO}=10$\,${\rm cm}^{-1}$), respectively.
Without loss of generality, we assume the two sites
have the same dipole strength and set $\mu_1=\mu_2=\mu_{r}$
for the Condon part, and $\mu_1'/\mu_{r}=\mu_2'/\mu_{r}=0.1$
for the non-Condon part [cf.\ \Eq{mua}],
with $\mu_r$ being the reference dipole strength.

\subsection{Field-triggered excitation energy transfer dynamics}
\label{thsec3A}

\begin{figure}
\includegraphics[width=0.5\textwidth]{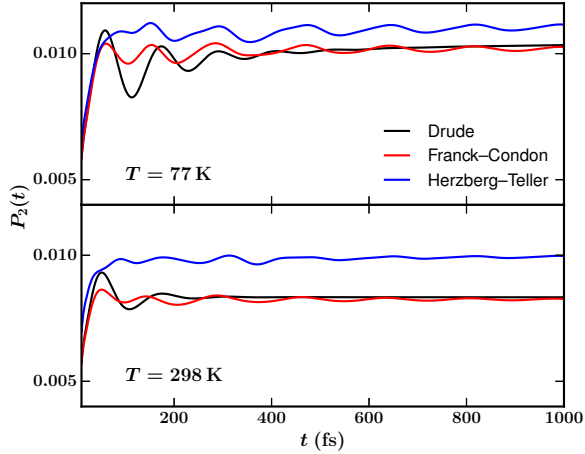}
\caption{Field-triggered EET dynamics for an excitonic dimer system at $77$ and $298$\,K:
Drude dissipation without vibronic coupling (black), 
with Franck--Condon coupling (red),
and with Herzberg--Teller coupling (blue).
The pulse is characterized by a Gaussian envelope $\mu_{r}\varepsilon(t)
=\frac{\theta}{\sqrt{2\pi}\sigma}\exp(-\frac{t^2}{2\sigma^2})\cos(\omega_c t)$,
with reference dipole strength $\mu_{r}$,
flipping angle $\theta=0.05\pi$,
temporal variance $\sigma=6$\,fs,
and carrier frequency $\omega_c=12400\,{\rm cm}^{-1}$.
}\label{fig4}
\end{figure}

In \Fig{fig4}, we report the field-triggered population dynamics
at three different conditions, including Drude dissipation
without vibronic coupling, with Franck--Condon coupling, 
and with Herzberg--Teller coupling.
It is observed that the time duration of the population oscillation is
largely elongated by both Condon and non-Condon vibronic couplings.
While the electronic decoherence with pure Drude dissipation lasts for about
hundreds of \,fs,
the vibronic decoherence reaches above $1$\,ps.
The oscillation is weakened
but still noticeable as the temperature increases.

We also find that the non-Condon vibronic coupling enhances
the quantum yield of the excitonic system. 
This effect becomes more obvious at the room temperature ($298$\, K), 
which indicates a stronger electronic-vibrational energy transfer therein.
In comparison, the Condon vibronic coupling does not vary
the final population significantly.
This finding highlights the dynamical interplay between 
excitation of the system and non-Condon polarization of the vibrational environment.

In the meanwhile, it is shown that
these two kinds of vibronic couplings result in similar population oscillation, 
indicating that the non-Condon vibronic coupling
retains the electronic coherence of the reduced system.
On the other hand, the total quantum coherence,
including contributions from both reduced system and polarized vibrational environment, is enhanced,
as discussed in the following section of the 2D spectroscopy simulations.

\subsection{Coherent two dimensional spectroscopy}
\label{thsec3B}

\begin{figure}
\includegraphics[width=0.5\textwidth]{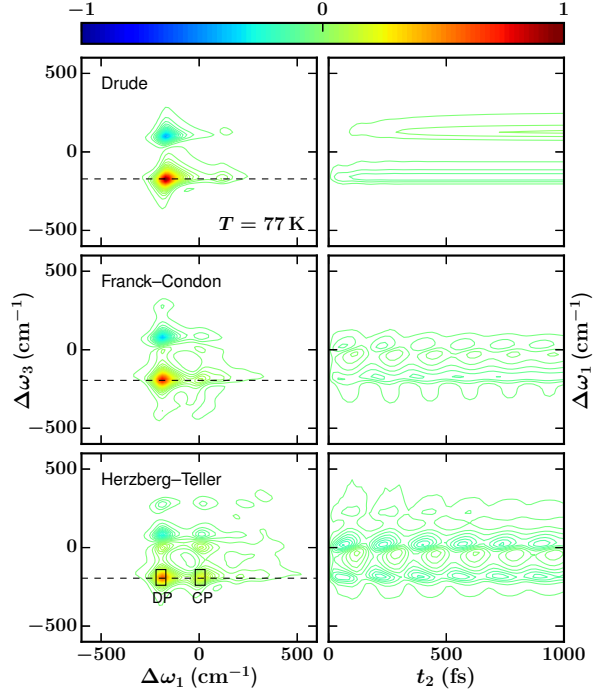}
\caption{
Two dimensional spectroscopy of the dimer system 
for Drude dissipation without vibronic coupling (top),
with Franck--Condon coupling (middle) 
and with Herzberg-Teller coupling (bottom), at $77$\,K. 
The left panels show the 2D signals at $t_2=0$,
while the right panels show the $t_2$ dynamics of 
the signals along the dashed horizontal line.
We set $\Delta\omega_k=\omega_k-(\epsilon_1+\epsilon_2)/2$ for $k=1$ and $3$.
All the results are scaled by the maximum of the pure Drude results,
and in the right panels, 
the initial values at $t_2=0$ have been subtracted.
}\label{fig5}
\end{figure}

\begin{figure}
\includegraphics[width=0.5\textwidth]{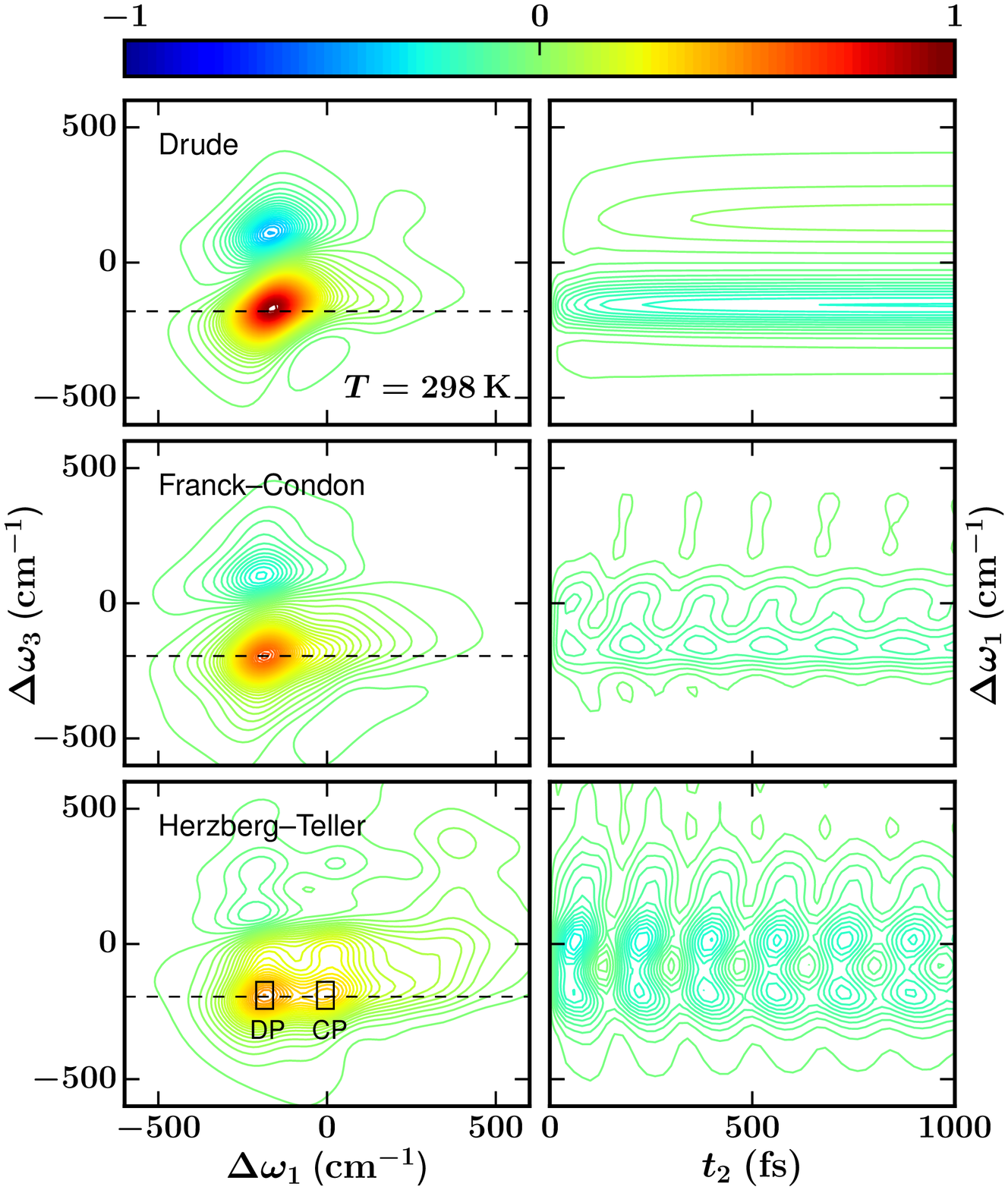}
\caption{
Two dimensional spectroscopy of the excitonic dimer in \Fig{fig5}
at $298$\,K.
All the results are scaled by the maximum of the pure Drude results.
}\label{fig6}
\end{figure}

To investigate the dynamical electronic-vibrational interaction
and the involving Condon and non-Condon effects,
we further simulate the 2D spectroscopy for the dimer system. 
2D coherent spectroscopy is a four-wave-mixing technique.\cite{Muk95,Muk042073,Muk091207}
The multi-chromophore system is excited by three time-ordered laser pulses (with
wavevectors ${\bf k}_1$ , ${\bf k}_2$ and ${\bf k}_3$) separated by time interval $t_1$ and $t_2$,
and after a time-delay $t_3$, the emitted signal is measured along the phase matching direction
${\bf k}_s=\pm {\bf k}_3\pm {\bf k}_2\pm {\bf k}_1$.
In this work, we simulate the pump-probe signal
that consists of rephasing ${\bf k}_{\rm I}={\bf k}_3+{\bf k}_2-{\bf k}_1$
and non-rephasing ${\bf k}_{\rm II}={\bf k}_3-{\bf k}_2+{\bf k}_1$ configurations.\cite{Muk95,Muk042073,Muk091207}
For simplicity, we adopt the rotating wave approximation and
the impulsive limit. \cite{Muk95,Muk042073,Muk091207}

The rephasing $S_{{\bf k}_{\rm I}}$ and non-rephasing 
$S_{{\bf k}_{\rm II}}$ signals are given by \cite{Muk95,Xu11497}
\begin{align}\label{S12}
  S_{{\bf k}_{\rm I/II}}(\omega_3,t_2,\omega_1)&= {\rm Re}
   \int_0^\infty\!\!\d t_3\!\!\int_0^\infty\!\!\d t_1
   e^{i(\omega_3t_3\mp\omega_1t_1)} \nl & \qquad \ \
  \times R_{{\bf k}_{\rm I/II}}(t_3,t_2,t_1).
\end{align}
and related respectively to the nonlinear optical correlation functions
\be\label{Rk1k2}
\begin{split}
  &R_{{\bf k}_{\rm I}}=R_2+R_3+R_5 \qquad \text{\ \ \ (rephasing)}\, ,
\\
 &R_{{\bf k}_{\rm II}}=R_1+R_4+R_6 \qquad \text{(non-rephasing)}\, .
\end{split}
\ee
In \Eq{Rk1k2}, the six Liouville-space pathways 
are expressed as\cite{Xu11497,Zha16CP}
\be\label{Rexpr}
\begin{split}
R_1&=
   \lla \hat{V}_{ge}(t_3)|\rightact{V}_{\!eg}{\cal G}_{ee}(t_2) \rightact{V}_{\!ge}
   {\cal G}_{eg}(t_1)  \leftact{V}_{\!eg}|{\rho}^{\rm eq}_{\T,gg}\rra,
\\
R_2&=
   \lla \hat{V}_{ge}(t_3)|\rightact{V}_{\!eg} {\cal G}_{ee}(t_2)  \leftact{V}_{\!eg}
   {\cal G}_{ge}(t_1) \rightact{V}_{\!ge}|{\rho}^{\rm eq}_{\T,gg}\rra,
\\
R_3&=
   \lla \hat{V}_{ge}(t_3)|\leftact{V}_{\!eg} {\cal G}_{gg}(t_2) \rightact{V}_{\!eg}
   {\cal G}_{ge}(t_1) \rightact{V}_{\!ge}|{\rho}^{\rm eq}_{\T,gg}\rra,
\\
R_4&=
   \lla \hat{V}_{ge}(t_3)|\leftact{V}_{\!eg} {\cal G}_{gg}(t_2)  \leftact{V}_{\!ge}
   {\cal G}_{eg}(t_1)  \leftact{V}_{\!eg}|{\rho}^{\rm eq}_{\T,gg}\rra,
\\
R_5&= -
   \lla \hat{V}_{ef}(t_3)|\leftact{V}_{\!fe} {\cal G}_{ee}(t_2)  \leftact{V}_{\!eg}
   {\cal G}_{ge}(t_1) \rightact{V}_{\!ge}|{\rho}^{\rm eq}_{\T,gg}\rra,
\\
R_6&= -
   \lla \hat{V}_{ef}(t_3)|\leftact{V}_{\!fe} {\cal G}_{ee}(t_2) \rightact{V}_{\!ge}
   {\cal G}_{eg}(t_1)  \leftact{V}_{\!eg}|{\rho}^{\rm eq}_{\T,gg}\rra.
\end{split}
\ee
Here, all the involving operators and the propagator ${\cal G}(t)\equiv e^{-i{\cal L}_{\T}t}$ are in their block forms
subscripted by ``$uv$'', where $u,v\in\{g,e,f\}$, with $|g\ra$, $|e\ra$ and $|f\ra$ specifying
the ground, singly-excited and doubly-excited electronic manifolds.\cite{Xu11497}
The efficient evaluation of the DEOM-space equivalence of \Eq{Rexpr}
follows Ref.\ \onlinecite{Zha16CP} and \onlinecite{Xu11497}.
In our simulations, all parameters 
remain the same as those in \Fig{fig4}.

Figure \ref{fig5} demonstrates the simulated pump-probe spectroscopy at $77$\,K 
under three circumstances, i.e. the pure Drude dissipation without vibronic coupling, 
the one with Franck--Condon coupling, and the one with Herzberg--Teller coupling. 
In the presence of the vibronic coupling, vibrational peaks on both diagonal
and off-diagonal are observed, while the exciton-exciton off-diagonal peak is concealed.
In particular, an electronic-vibrational cross peak (CP) appears 
right to the excitonic diagonal peak (DP) with the frequency difference around $\omega_{\BO}$, 
as observed in the middle- and bottom-left panels of \Fig{fig5}.
It suggests that this CP measures the cross correlation between
the first and the zeroth vibrational levels within the excited electronic state.
With the non-Condon vibronic coupling,
the above vibrational signals are more prominent.
The right panels show the $t_2$ time-evolution of the signals along the dashed horizontal line,
covering both the excitonic DP and the electronic-vibrational CP.
With the vibronic coupling, the duration of the quantum beats is elongated obviously. As shown in the 
middle- and bottom-right panels, the vibronic CPs oscillate for at least 1\,ps with both 
Condon and non-Condon vibronic couplings. 
For DPs, however, the oscillation with non-Condon coupling is more persistent than that with Condon coupling. 
Moreover, the non-Condon coupling results in larger oscillation amplitude. 
Evidently, the vibronic coupling facilitates the quantum coherence in the EET dynamics.
In particular, the non-Condon transition greatly enhances the excitation of
higher vibrational levels in the excited state.
The subsequent vibrational energy relaxation preserves
the long-lived coherence and amplifies the oscillation amplitudes in 2D spectroscopy.

\begin{figure}
\includegraphics[width=0.5\textwidth]{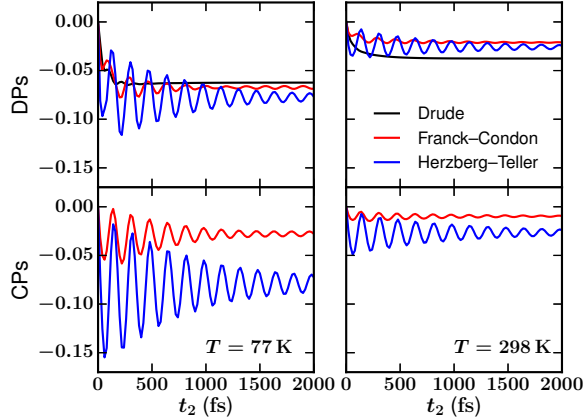}
\caption{
Amplitudes of the DPs and CPs along time interval $t_2$, 
extracted from the 2D spectroscopy with Franck--Condon coupling (black) and 
the Herzberg--Teller coupling (red) at $77$ and $298$\,K.
All curves are scaled by the maximum of pure Drude results at $77$\,K
and shifted vertically to make the initial values $0$.
}\label{fig7}
\end{figure}

In comparison with the results at $77$\,K, 
the 2D signals at $298$\,K are much broader, 
and the electronic and vibrational excitation peaks
are not well separated, see \Fig{fig6}.
However, in the right panels, we observe similar dynamics of the excitonic DPs 
and electronic-vibrational CPs as those at $77$\,K.
Apparently, the vibrational energy relaxation at $298$\,K is more
prominent than that at $77$\,K, see the middle-right and bottom-right panels
in \Fig{fig5} and \Fig{fig6} for comparison.

In \Fig{fig7}, we further compare the oscillation amplitudes at the cryogenic 
and room temperatures of the excitonic DPs and vibronic CPs along $t_2$,
with all curves being scaled by the maximum of the pump-probe spectroscopy with pure Drude dissipation at 
cryogenic temperature ($77$\,K).
In both the Condon and non-Condon couplings, the quantum beats of the DPs and CPs last for more than $2$\,ps , and their 
amplitudes are still noticeable at room temperauture ($298$\,K),
about one-third of those at $77$\,K.
In contrast, the quantum beats from purely electronic transitions 
are very short-lived, i.e., its coherence only lasts for 300 fs at $77$\,K and becomes even more transient at $298$\,K. 
Moreover, the non-Condon vibronic coupling
further intensifies the vibrational oscillation amplitudes dramatically at both temperatures,
in comparison with the Condon results.

Clearly, 2D spectroscopy detects both the 
excitonic system and the polarized vibrational environment.
Our simulation results indicate that the vibronic coupling 
plays a dominant role in preserving the long-lived quantum beats,
whereas the purely electronic coherence has minor contribution.
This persistent quantum coherence is present at both cryogenic and room temperatures.
Moreover, in comparison with the Condon counterpart, 
the non-Condon vibronic coupling greatly enhances the 
excitation of the higher vibrational levels
in the excited electronic state, and the subsequent vibrational relaxation 
leads to the stronger quantum beats,
in the $t_2$-resolved pump-probe spectroscopy.

\section{Conclusion}
\label{thconc}

In this work, we investigate the non-Condon virbonic coupling effects
on the quantum coherence of the EET dynamics,
via the non-perturbative DEOM simulations on excitonic model systems.
Evidently, the DEOM is a unique theory
in studying both the reduced system and the hybrid bath dynamics
as well as their correlated quantum entanglement.
Our simulations on field-triggered EET dynamics
and coherent 2D spectroscopy show that the vibronic coupling
elongates the quantum beats at both cryogenic and room temperatures.
In particular, the non-Condon vibronic coupling facilitates
the electronic-vibrational energy transfer
and dramatically enhances the total system-and-bath quantum coherence.
This work provides an insightful investigation on the origin of the
quantum coherence, and may enrich the
interpretation of the experimental nonlinear spectroscopy.

\section*{Acknowledgements}
The support from 
the Natural Science Foundation of China (Nos.\ 21373191 and 21633006),
the Fundamental Research Funds for the Central Universities (No.\ 2030020028),
and the Ministry of Science and Technology (No.\ 2016YFA0400900)
is gratefully acknowledged.


\begin{thebibliography}{10}

\bibitem{Eng07782}
G.~S. Engel et~al.,
\newblock Nature {\bf 446}, 782 (2007).

\bibitem{Che084254}
Y.-C. Cheng and G.~R. Flemming,
\newblock J. Phys. Chem. B {\bf 112}, 4254 (2008).

\bibitem{Pan1012766}
G.~Panitchayangkoon et~al.,
\newblock Proc. Natl. Acad. Sci. USA {\bf 107}, 12766 (2010).

\bibitem{Ple13235102}
M.~B. Plenio, J.~Almeida, and S.~F. Huelga,
\newblock J. Chem. Phys. {\bf 139}, 235102 (2013).

\bibitem{Tur111904}
D.~B. Turner, K.~E. Wilk, P.~M.~G. Curmi, and G.~D. Scholes,
\newblock J. Phys. Chem. Lett. {\bf 2}, 1904 (2011).

\bibitem{Tem1412865}
R.~Tempelaar, T.~L.~C. Jansen, and J.~Knoester,
\newblock J. Phys. Chem. B {\bf 118}, 12865 (2014).

\bibitem{But14034306}
V.~Butkus, L.~Valkunas, and D.~Abramavicius,
\newblock J. Chem. Phys. {\bf 140}, 034306 (2014).

\bibitem{Cam1595}
F.~V.~A. Camargo, H.~L. Anderson, S.~R. Meech, and I.~A. Heisler,
\newblock J. Phys. Chem. A {\bf 119}, 95 (2015).

\bibitem{Bal1519491}
V.~Balevi{\v c}ius et~al.,
\newblock Phys. Chem. Chem. Phys. {\bf 17}, 19491 (2015).

\bibitem{Man121497}
T.~Man{\v c}al et~al.,
\newblock J. Phys. Chem. Lett. {\bf 3}, 1497 (2012).

\bibitem{Per14404}
V.~Perl{\'i}k, C.~Lincoln, F.~{\v S}anda, and J.~Hauer,
\newblock J. Phys. Chem. Lett. {\bf 5}, 404 (2014).

\bibitem{Fuj15212403}
Y.~Fujihashi, G.~R. Fleming, and A.~Ishizaki,
\newblock J. Chem. Phys. {\bf 142}, 212403 (2015).

\bibitem{Mon15065101}
D.~M. Monahan, L.~Whaley-Mayda, A.~Ishizaki, and G.~R. Fleming,
\newblock J. Chem. Phys. {\bf 143}, 065101 (2015).

\bibitem{Ish082219}
K.~Ishii, S.~Takeuchi, and T.~Tahara,
\newblock J. Phys. Chem. A {\bf 112}, 2219 (2008).

\bibitem{Kam0010637}
P.~Kambhampati, D.~H. Son, T.~W. Kee, and P.~F. Barbara,
\newblock J. Phys. Chem. A {\bf 104}, 10637 (2000).

\bibitem{Wom12154016}
J.~M. Womick, B.~A. West, N.~F. Scherer, and A.~M. Moran,
\newblock J. Phys. B: At. Mol. Opt. Phys. {\bf 45}, 154016 (2012).

\bibitem{Tan89101}
Y.~Tanimura and R.~Kubo,
\newblock J. Phys. Soc. Jpn. {\bf 58}, 101 (1989).

\bibitem{Yan04216}
Y.~A. Yan, F.~Yang, Y.~Liu, and J.~S. Shao,
\newblock Chem. Phys. Lett. {\bf 395}, 216 (2004).

\bibitem{Ish053131}
A.~Ishizaki and Y.~Tanimura,
\newblock J. Phys. Soc. Jpn. {\bf 74}, 3131 (2005).

\bibitem{Tan06082001}
Y.~Tanimura,
\newblock J. Phys. Soc. Jpn. {\bf 75}, 082001 (2006).

\bibitem{Tan14044114}
Y.~Tanimura,
\newblock J. Chem. Phys. {\bf 141}, 044114 (2014).

\bibitem{Tan15144110}
Y.~Tanimura,
\newblock J. Chem. Phys. {\bf 142}, 144110 (2015).

\bibitem{Xu05041103}
R.~X. Xu, P.~Cui, X.~Q. Li, Y.~Mo, and Y.~J. Yan,
\newblock J. Chem. Phys. {\bf 122}, 041103 (2005).

\bibitem{Xu07031107}
R.~X. Xu and Y.~J. Yan,
\newblock Phys. Rev. E {\bf 75}, 031107 (2007).

\bibitem{Zhe121129}
X.~Zheng et~al.,
\newblock Prog. Chem. {\bf 24}, 1129 (2012),
\newblock http://www.progchem.ac.cn/EN/abstract/abstract10858.shtml.

\bibitem{Jin08234703}
J.~S. Jin, X.~Zheng, and Y.~J. Yan,
\newblock J. Chem. Phys. {\bf 128}, 234703 (2008).

\bibitem{Che11194508}
L.~P. Chen, R.~H. Zheng, Y.~Y. Jing, and Q.~Shi,
\newblock J. Chem. Phys. {\bf 134}, 194508 (2011).

\bibitem{Yan14054105}
Y.~J. Yan,
\newblock J. Chem. Phys. {\bf 140}, 054105 (2014).

\bibitem{Zha15024112}
H.~D. Zhang, R.~X. Xu, X.~Zheng, and Y.~J. Yan,
\newblock J. Chem. Phys. {\bf 142}, 024112 (2015).

\bibitem{Jin15234108}
J.~S. Jin, S.~K. Wang, X.~Zheng, and Y.~J. Yan,
\newblock J. Chem. Phys. {\bf 142}, 234108 (2015).

\bibitem{Xu09214111}
R.~X. Xu, B.~L. Tian, J.~Xu, Q.~Shi, and Y.~J. Yan,
\newblock J. Chem. Phys. {\bf 131}, 214111 (2009).

\bibitem{Zhu115678}
K.~B. Zhu, R.~X. Xu, H.~Y. Zhang, J.~Hu, and Y.~J. Yan,
\newblock J. Phys. Chem. B {\bf 115}, 5678 (2011).

\bibitem{Din11164107}
J.~J. Ding, J.~Xu, J.~Hu, R.~X. Xu, and Y.~J. Yan,
\newblock J. Chem. Phys. {\bf 135}, 164107 (2011).

\bibitem{Din12224103}
J.~J. Ding, R.~X. Xu, and Y.~J. Yan,
\newblock J. Chem. Phys. {\bf 136}, 224103 (2012).

\bibitem{Zha15214112}
H.~D. Zhang and Y.~J. Yan,
\newblock J. Chem. Phys. {\bf 143}, 214112 (2015).

\bibitem{Xu151816}
R.~X. Xu, H.~D. Zhang, X.~Zheng, and Y.~J. Yan,
\newblock Sci. China Chem. {\bf 58}, 1816 (2015),
\newblock Special Issue: Lemin Li Festschrift.

\bibitem{Zha16CP}
H.~D. Zhang, Q.~Qiao, R.~X. Xu, and Y.~J. Yan,
\newblock Chem. Phys. (2016) doi: 10.1016/j.chemphys.2016.07.005.

\bibitem{Orl734513}
G.~Orlandi and W.~Siebrand,
\newblock J. Chem. Phys. {\bf 58}, 4513 (1973).

\bibitem{Lin743802}
S.~H. Lin and H.~Eyring,
\newblock Proc. Nat. Acad. Sci. USA {\bf 71}, 3802 (1974).

\bibitem{Tan932263}
Y.~Tanimura and S.~Mukamel,
\newblock J. Opt. Soc. Am. B {\bf 10}, 2263 (1993).

\bibitem{Tan93118}
Y.~Tanimura and S.~Mukamel,
\newblock Phys. Rev. E {\bf 47}, 118 (1993).

\bibitem{Tho012991}
M.~Thoss, H.~B. Wang, and W.~H. Miller,
\newblock J. Chem. Phys. {\bf 115}, 2991 (2001).

\bibitem{Wei08}
U.~Weiss,
\newblock {\em Quantum Dissipative Systems},
\newblock World Scientific, Singapore, 2008,
\newblock 3rd ed. Series in Modern Condensed Matter Physics, Vol.\ 13.

\bibitem{Yan05187}
Y.~J. Yan and R.~X. Xu,
\newblock Annu. Rev. Phys. Chem. {\bf 56}, 187 (2005).

\bibitem{Muk95}
S.~Mukamel,
\newblock {\em The Principles of Nonlinear Optical Spectroscopy},
\newblock Oxford University Press, New York, 1995.

\bibitem{Fey63118}
R.~P. Feynman and F.~L. \mbox{Vernon, Jr.},
\newblock Ann. Phys. {\bf 24}, 118 (1963).

\bibitem{Xu11497}
J.~Xu, R.~X. Xu, D.~Abramavicius, H.~D. Zhang, and Y.~J. Yan,
\newblock Chin. J. Chem. Phys. {\bf 24}, 497 (2011).

\bibitem{Xu13024106}
J.~Xu, H.~D. Zhang, R.~X. Xu, and Y.~J. Yan,
\newblock J. Chem. Phys. {\bf 138}, 024106 (2013).

\bibitem{Muk042073}
S.~Mukamel and D.~Abramavicius,
\newblock Chem. Rev. {\bf 104}, 2073 (2004).

\bibitem{Muk091207}
S.~Mukamel, Y.~Tanimura, and P.~Hamm,
\newblock Special Issue of Acc. Chem. Res. {\bf 42}, 1207 (2009).

\end{thebibliography}

\end{document}